\documentclass[twocolumn,showpacs,prl]{revtex4}% Physical Review Letters

\usepackage{graphicx}% Include figure files
\usepackage{dcolumn}% Align table columns on decimal point
\usepackage{bm}% bold math
\usepackage{color}

\begin{document}

%\preprint{???}

\title{Dynamics of polymers in elongational flow investigated by inelastic neutron scattering}

\author{Maikel~C.~Rheinst\"adter$^{1,2}$}\email{RheinstadterM@missouri.edu}
\author{Rainer Sattler$^3$}
\author{Wolfgang H\"au{\ss}ler$^4$}
\author{Christian Wagner$^3$}\email{c.wagner@mx.uni-saarlande.de}

\affiliation{$^1$Department of Physics and Astronomy, University of
Missouri Columbia, Columbia, MO 65211, U.S.A.\\
$^2$Institut Laue-Langevin, 6 rue Jules Horowitz, BP 156, 38042
Grenoble Cedex 9, France\\
$^3$Technische Physik, Universit\"at des Saarlandes, Geb.E2 6
Postfach 151150, 66041 Saarbr\"ucken,
Germany\\
$^4$FRM-II \& E21, Technische Universit\"at M\"unchen,
Lichtenbergstrasse 1, 85747 Garching, Germany}

\date{\today}% It is always \today, today,
             %  but any date may be explicitly specified

%------------------------------------------------------------------------------------
\begin{abstract}
The nanoscale fluctuation dynamics of semi dilute high molecular
weight polymer solutions of Polyethylenoxide (PEO) in D$_2$O under
non-equilibrium flow conditions were studied by the neutron
spin-echo
technique. %The application of this complex technology provides an
%unique access to the dynamics on the nano meter scale.
The sample cell was in contraction flow geometry and provided a
pressure driven flow with a high elongational component that
stretched the polymers most efficiently. The experiments suggest
that the mobility on the scale of a few monomers, comparable to the
Kuhn segment length, becomes highly anisotropic and is enhanced
perpendicular to the flow direction. Diffraction data show a weak
structural correlation along the chains on a length scale of about
17~\AA, which might be related to the Kuhn length in this system.

%The perpendicular component, $\tau_{\bot}$, can be described by an
%analytical model, which predicts faster relaxation times with
%increasing flow rate $\dot\epsilon$, while the predicted slowing of
%the parallel component, $\tau_{\|}$, could not be observed.
\end{abstract}

% 83.85.Hf X-ray and neutron scattering
% 83.10.Mj Molecular dynamics, Brownian dynamics

%\keywords{Suggested keywords}%Use showkeys class option if keyword
                              %display desired
\maketitle

%------------------------------------------------------------------------------------
%\section{Introduction\label{Introduction}}
\begin{figure}
\centering
\resizebox{0.28\columnwidth}{!}{\rotatebox{0}{\includegraphics{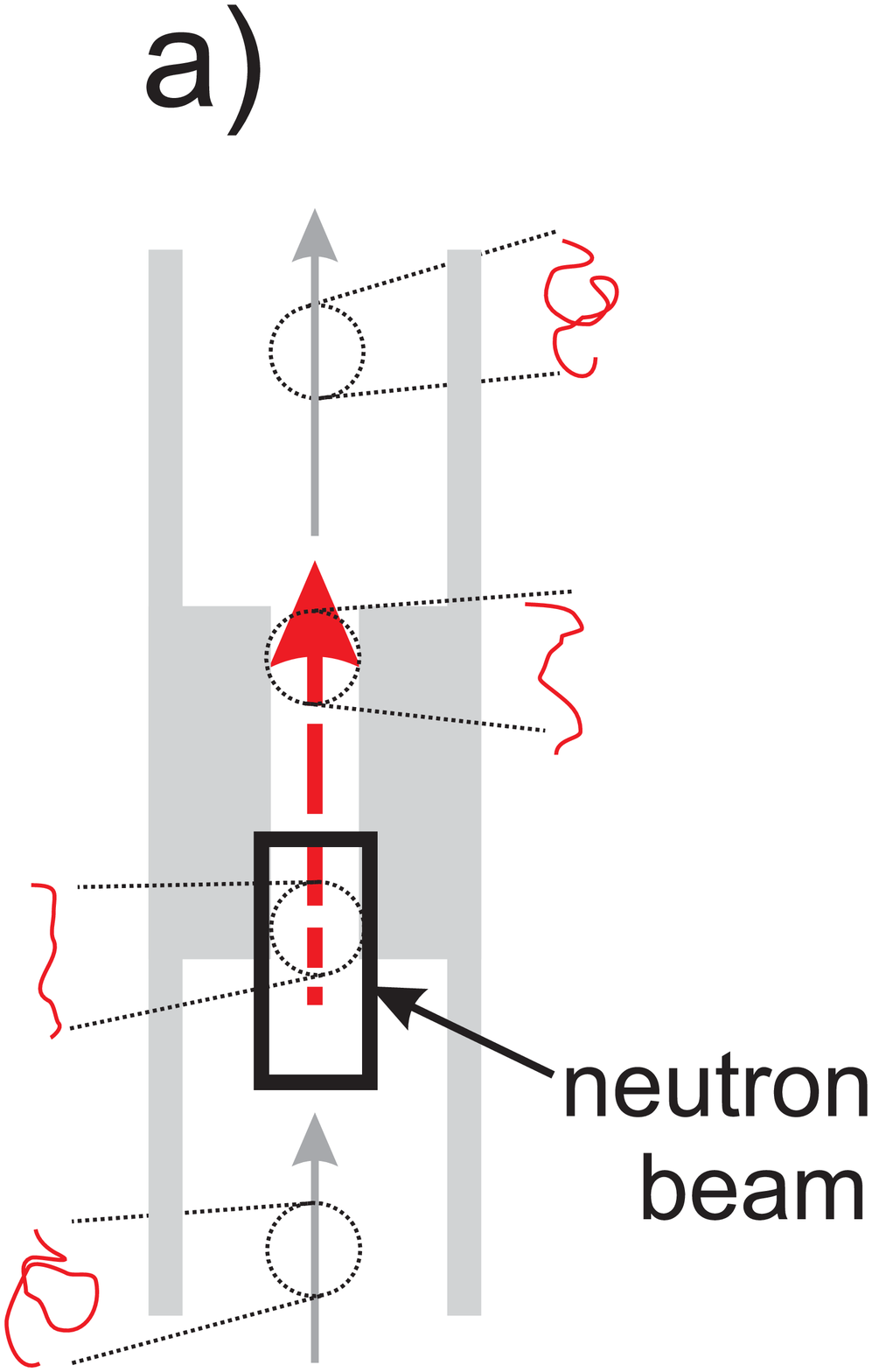}}}
\resizebox{0.70\columnwidth}{!}{\rotatebox{0}{\includegraphics{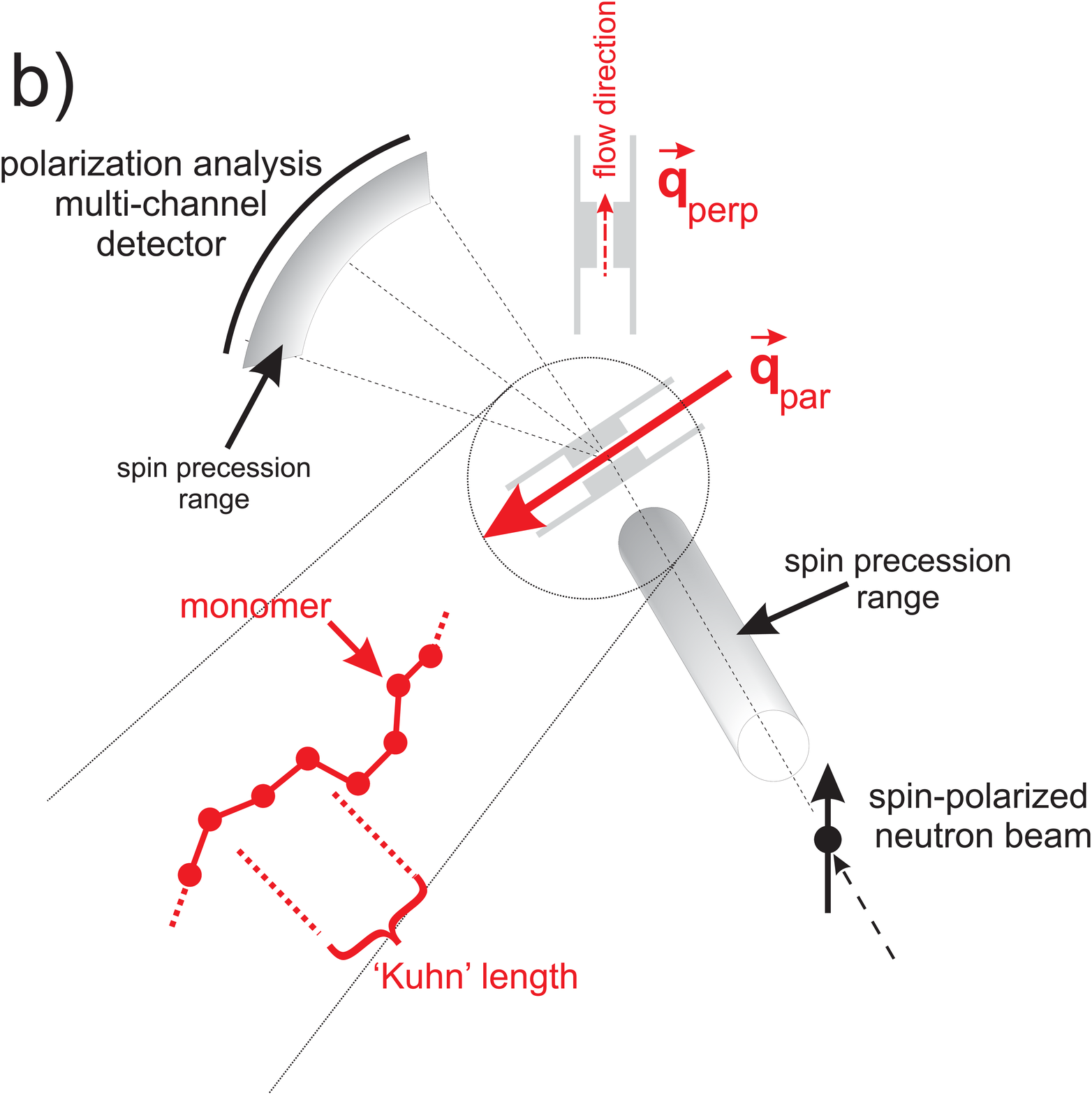}}}
\caption{(Color online) (a) Scheme of the flow cell. The flow is
nonlinear and polymers are stretched mostly shortly before entering
the contraction. With increasing flow rate, the polymer is more and
more stretched in the elongational flow. Size and position of the
neutron beam is marked. (b) Sketch of the experimental set-up. The
flow cell can be aligned parallel or perpendicular to the scattering
plane.}\label{PolymerSketch.eps}
\end{figure}
The study of the flow behavior of polymer solutions is both of
scientific and applied interest. Polymer solutions are present in
our daily life: Most biological liquids are polymer solutions, food
often contains polymers to control the flow behavior and they are
also used in industrial applications, such as lubrication.
Challenging from a scientific point of view is the relation of a
microscopic system with an almost infinite number of degrees of
freedom, namely the polymer chains, and the macroscopic flow
properties, preferably described with a continuum theoretical
approach.  The way polymer containing liquids flow differs
significantly from the behavior of simple liquids. Many new
phenomena, such as the shear rate dependent viscosity or the normal
stress, arise due to the elastic behavior of the microscopic
constituents, i.e., the polymer chains. In the last decade the
understanding of the equilibrium dynamics of polymers in solutions
made significant progress, both due to modern experimental
techniques, such as single molecule fluorescence spectroscopy
\cite{Perkins1997,Schwille2006}, but also elastic and inelastic
neutron scattering, and new statistical concepts that lead to a
better theoretical description. A major difficulty is the large
spectrum of scales involved, ranging from meters and hours in
polymer processing down to Angstroems and nanoseconds on the monomer
scale. Broadly accepted theories based on bead-rod or bead-spring
models \cite{Bird87} describe the polymer molecule as a chain of
segments with certain degrees of freedom and elasticity.

It is still not well understood how an applied flow field affects
structure and dynamics of polymers on the nanometer scale, i.e., on
the scale of a single {\em Kuhn segment}. The Kuhn segment is the
smallest statistical unit that gives an estimate of the length scale
on which orientational information is lost (the "effective" segment
length in the ideal chain model). Any flow can be divided into a
{\em rotational} and an {\em elongational} part. While the rotation
has no deforming effect on the polymers, the elongational flow
beyond a critical rate \cite{Perkins1997, Larson2003} stretches
polymers at maximum \cite{Wagner2003}. The associated entropic
elastic forces of the polymers give rise to elongational viscosities
that might be several orders of magnitude larger than those measured
in shear flow \cite{Anna2001,Stelter2002}. Flow with a strong
elongational component arises typically in many relevant flow
situations, such as in fiber spinning, but also in any extrusion
process or in contraction flow. Only recently, by using small angle
neutron scattering (SANS) it was shown that such flow geometry
affects the structure, i.e., the radius of gyration of polymers in a
polymer melt \cite{Bent2003}. The question that we address here is
how an elongational flow affects the {\em dynamics} of the
statistical segments of the polymer. By using the neutron spin echo
technique (NSE), we were able to investigate the anisotropic
fluctuation dynamics of the chain segments parallel and
perpendicular to an elongational flow. NSE accesses much smaller
length scales as compared to dynamical light scattering (DLS) and
x-ray photon correlation spectroscopy (XPCS) and allows to study
structure and dynamics down to the monomer scale of the polymers in
solution or melt.

%------------------------------------------------------------------------------------
%\section{Experimental\label{Experimental}}
We investigated polymer dynamics of a semidiluted Polyethylenoxide
(PEO) solution in deuterated water (D$_2$O) to increase the
scattering contrast and signal. PEO is known for its large
flexibility and its effect on elongational flow is at maximum for
all polymeric systems \cite{Perkins1997,Wagner2003}. Solutions of
protonated PEO of M$_{W}$=4~Mio~amu at a concentration of $0.3$
weight percent in D$_{2}$O were prepared. Complementary rheological
data in shear (MARS, ThermoScientific, Karlsruhe) and elongational
(CABER, ThermoScientific, Karlsruhe) experiments did not show
differences to solutions in H$_2$O. The overlap equilibrium
concentration at which polymers start to interact with each other
can be estimated to $c^*=0.07$ weight percent and we thus stayed in
the semi diluted regime. % The solvability of PEO at concentrations
%higher than $0.3$ weight percent was poor and a higher tendency to a
%gel like transition or flocculation was observed.
The characteristic ratio $c_\infty = 5.5$ gives a measure on how
many monomers build up a segment of the freely rotating chain model.
It can be estimated from rheological data, together with the Kuhn
segment length of $l\simeq 10$ \AA\ in this system and the data are
tabulated in the literature \cite{Brandrup1999}.

Figure~\ref{PolymerSketch.eps} (a) shows the schematic of the flow
cell in which contraction flow was generated \cite{Bent2003}. We
used aluminum cells in $4:1$ contraction ratio. The three different
segments of the flow cell were 10~cm long, each, and the width
varied from  4~cm to 1~cm with a constant inner thickness of 5~mm.
As depicted in Fig.~\ref{PolymerSketch.eps} (a), we expect a strong
anisotropy in local relaxation processes when the polymer is
(depending on the flow rate) more and more stretched out in the
flow. Dry compressed air was used to push the liquid from a
container through the contraction towards a second container placed
on a precision scale to control the flow rates to 60~ml/min and
100~ml/min, yielding to maximum elongational rates of
$\dot\epsilon_1\approx 4\cdot s^{-1}$ (Flow 1) and
$\dot\epsilon_2\approx 6.7\cdot s^{-1}$ (Flow 2). Complementary
elongational rheological data show that these elongational rates and
the transfer time through the contraction area are high enough to
stretch the polymers but still well below the onset of elastic
instabilities \cite{Rodd2007}. The back flow to the reservoir was
generated by a flexible-tube pump. The inevitable strong shear rates
in the tubes lead to an overall aging of the polymer solutions,
which has been monitored by measuring the elongation viscosity.
While the viscosity decreased from 10~Pa$\cdot$s to 5~Pa$\cdot$s
within the first 12~h, no further significant degradation was found
for another 48 hours of pumping.
% The
%elongational viscosity is very sensitive to the length of the
%polymer chains and we observed a diminution of the elongational
%viscosity from 10~Pa$\cdot$s to 5~Pa$\cdot$s after 12~h of pumping.
%No further significant degradation was found for another 48 hours of
%pumping. Data acquisition in the scattering experiments thus started
%after 12 hours of pumping and the samples were renewed after 48
%hours.

A spin-echo spectrometer turned out to be highly suited for the
inelastic experiments. The spin-echo technique offers extremely high
energy resolution from Larmor tagging the neutrons \cite{Mezei:1980}
and works in the time domain by measuring the intermediate
scattering function $S(q_{(\|,\perp)},t)$. For a quasielastic energy
response, assumed to have Lorentzian lineshape with half-width
$\Gamma$, the polarization will show a single exponential decay
$P_{NSE} = P_s \exp{(-\Gamma t)}$ in the time domain. Note that the
flow rates were in the order of millimeters per second and did
therefore not contaminate the intermediate scattering function by an
inelastic Doppler peak. Scattering of the functional polymer units
was measured using the multi angle spin-echo spectrometer IN11C,
equipped with 40 single counters, at the high flux neutron source of
the Institut Laue-Langevin in Grenoble, France, operated at a
wavelength of $\lambda=5.5$~\AA. $q$-values of 0.2 to
$0.8~$\AA$^{-1}$ corresponding to length scales of about 8 to
30~\AA\ (several times the monomer length scale) were measured
simultaneously, covering time scales from 4~ps to 1~ns. The
spectrometer can be used to measure elastic scattering in a
relatively coarse resolution, but simultaneously to the inelastic
runs, what delivers important information to interpret the dynamical
data. Figure~\ref{PolymerSketch.eps} (b) shows a sketch of the
experimental setup. By rotating the flow cell by 90$^{\circ}$
perpendicular to the neutron beam, the scattering vector ${\bf q}$
could be placed in the direction of the flow, $q_{\|}$, or
perpendicular to it, $q_{\perp}$. Because of the small scattering
signal, typical scans were performed by continuously integrating
over 24~h. Temperature was not controlled, but room temperature was
around 25~$^{\circ}$C and stable within $\pm$2 degrees.

\begin{figure}
\centering
\resizebox{1.00\columnwidth}{!}{\rotatebox{0}{\includegraphics{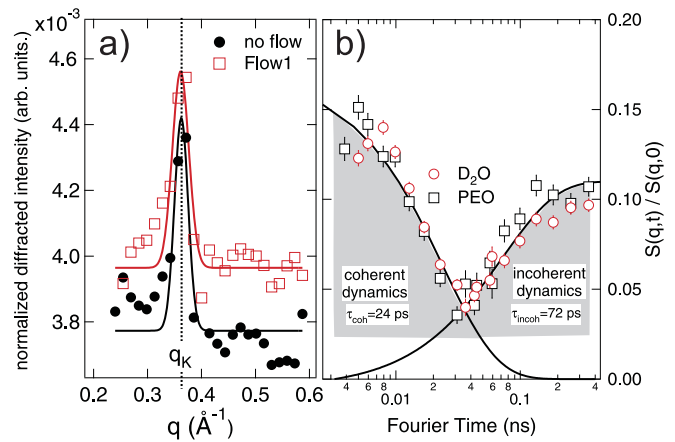}}}
%\resizebox{0.90\columnwidth}{!}{\rotatebox{0}{\includegraphics{Fig2_GraphDiffraction.eps}}}
\caption{(Color online) (a) In-situ diffraction of the PEO solution
at rest and with applied flow (Flow 1). The weak peak at
$q_K=0.363~$\AA$^{-1}$ ($2\pi/q_K=17.3~$\AA) most likely stems from
short range dynamical order within the polymer chains. Solid lines
are fits after Gaussian peak shapes. Curves have been shifted
vertically. (b) Intermediate scattering function integrated over $q$
values from $0.2~$\AA$^{-1}<q<0.8~$\AA$^{-1}$. Two steps
corresponding to coherent and incoherent scattering contributions
are visible.} \label{GraphDiffraction.eps}
\end{figure}
Figure~\ref{GraphDiffraction.eps} (a) depicts in-situ diffraction
data for the polymer solution at rest and with an applied flow (Flow
1). After integrating over 24 hours, the diffracted intensity showed
a very weak peak at a $q$-value of $q_K=0.363~$\AA$^{-1}$ or
($2\pi/q_K=17.3~$\AA), which most likely is not produced by an
internal polymer structure, but stems from short range dynamical
order within the polymer chains. The length scale of this ordering
is on the length scale of the Kuhn segments' length in PEO as
estimated from rheological data. The correlation does not depend on
the applied flow and is isotropic as the peak is observed in
$q_{\|}$ and $q_{\bot}$. Only on the length scale of this peak, the
dynamics of the polymer solution is different from that of the
solvent, as will be shown below. This finding is very common in
soft-matter, and also in liquids, where only around the correlation
peak (which is also defined by a short range dynamic order), long
living excitations are observed in the 'roton minimum' of the
corresponding dispersion curve.

Inelastic scans were taken for ${\bf q}$ parallel and perpendicular
to the flow direction for two different flow speeds. As a survey,
Figure~\ref{GraphDiffraction.eps} (b) shows results integrated over
the whole $q$-range for pure D$_2$O and PEO solution at rest. By
normalizing the data to a perfectly elastic scatterer (TiZr),
$S(q_{(\|,\perp)},t)/S(q_{(\|,\perp)},0)=1$ means purely elastic
scattering signal (no dynamics). The observed values of
$S(q_{(\|,\perp)},t)/S(q_{(\|,\perp)},0)\approx 0.15$ show that more
than 85\% of the dynamics is faster and outside of the time window
observed, most likely due to fast molecular motions of the carrier
liquid and internal polymer dynamics. The total scattering signal
shows a decay at small Fourier times and an increase at higher times
and can be fitted by two exponential steps:
$S(q_{(\|,\perp)},t)/S(q_{(\|,\perp)},0)^{total}=A_{coh}
\exp{(-t/\tau_{coh})}+A_{inc} \exp{(-t/\tau_{inc})}$. The coherent
contribution is dominant at smaller Fourier times $t\lesssim
0.04$~ns. The amplitude of the spin-incoherent scattering, which is
prominent for protons, $A_{inc}$, is negative because it involves a
spin-flip of the precessing neutron spin leading to an upward step
\cite{Richter:2005}. Diffusion and local internal dynamics, such as
vibrations and short range rotations, of water molecules and
polymers contribute to the incoherent scattering. Coherent
scattering probes correlated molecular motions, i.e., collective
water, but also correlated chain dynamics of the polymers (in the
range of length and time scales covered). The relaxation times
$\tau_{coh}$ and $\tau_{inc}$ for D$_2$O and PEO at rest were
determined from the fits and agree well with corresponding values
reported for H$_2$O \cite{Dore:1991}. Because of the low polymer
concentration, the total inelastic signal is strongly dominated by
coherent and incoherent scattering of the carrier liquid. Even
though D$_2$O was used to minimize the incoherent scattering of the
carrier liquid, it still completely covered a possible contribution
of the polymer chains to the total scattering above $t\gtrsim
0.04$~ns. So the analysis will focus on the faster times in the
following.

\begin{figure*}
\centering
\resizebox{0.31\textwidth}{!}{\rotatebox{0}{\includegraphics{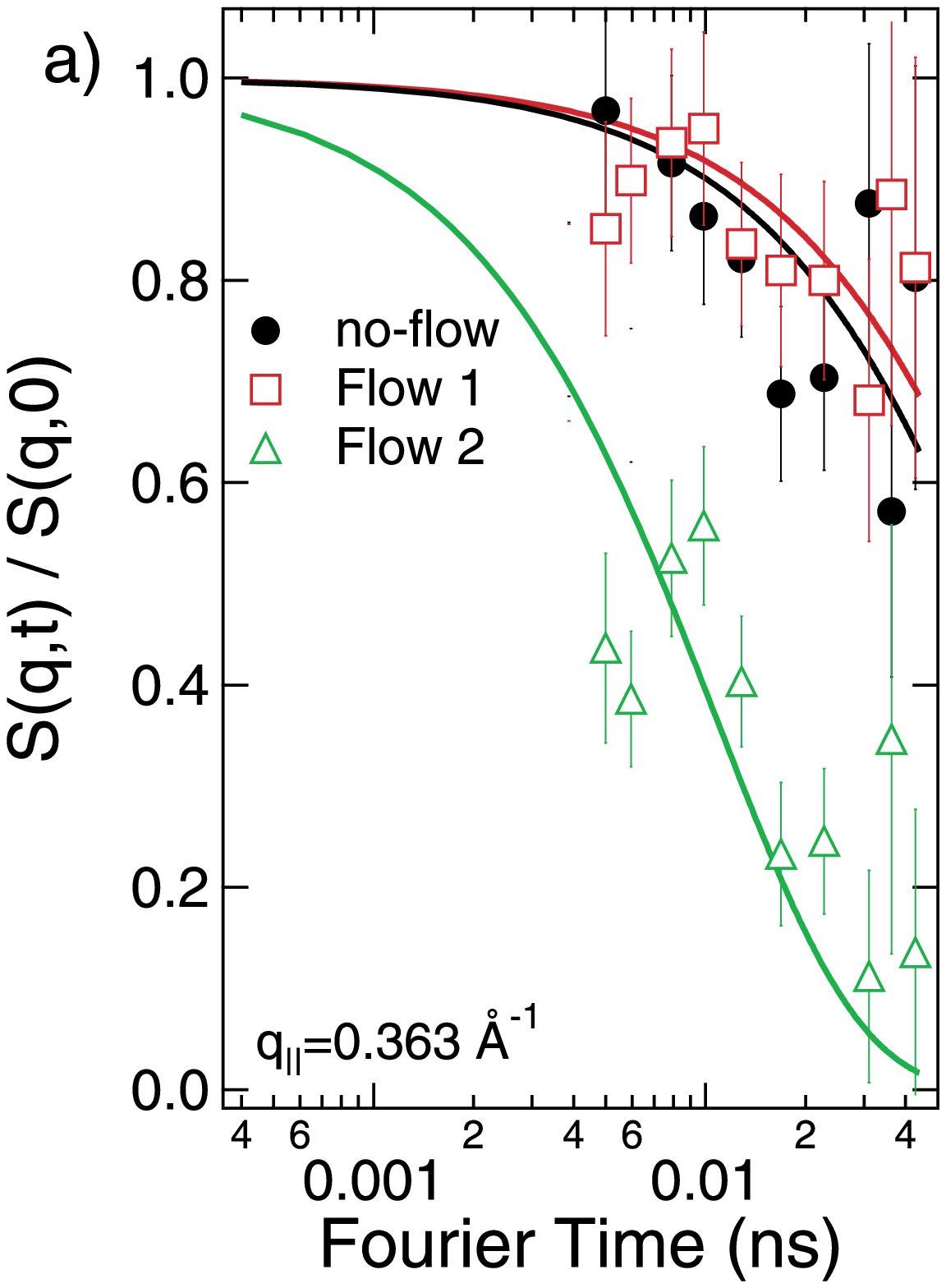}}}
\resizebox{0.31\textwidth}{!}{\rotatebox{0}{\includegraphics{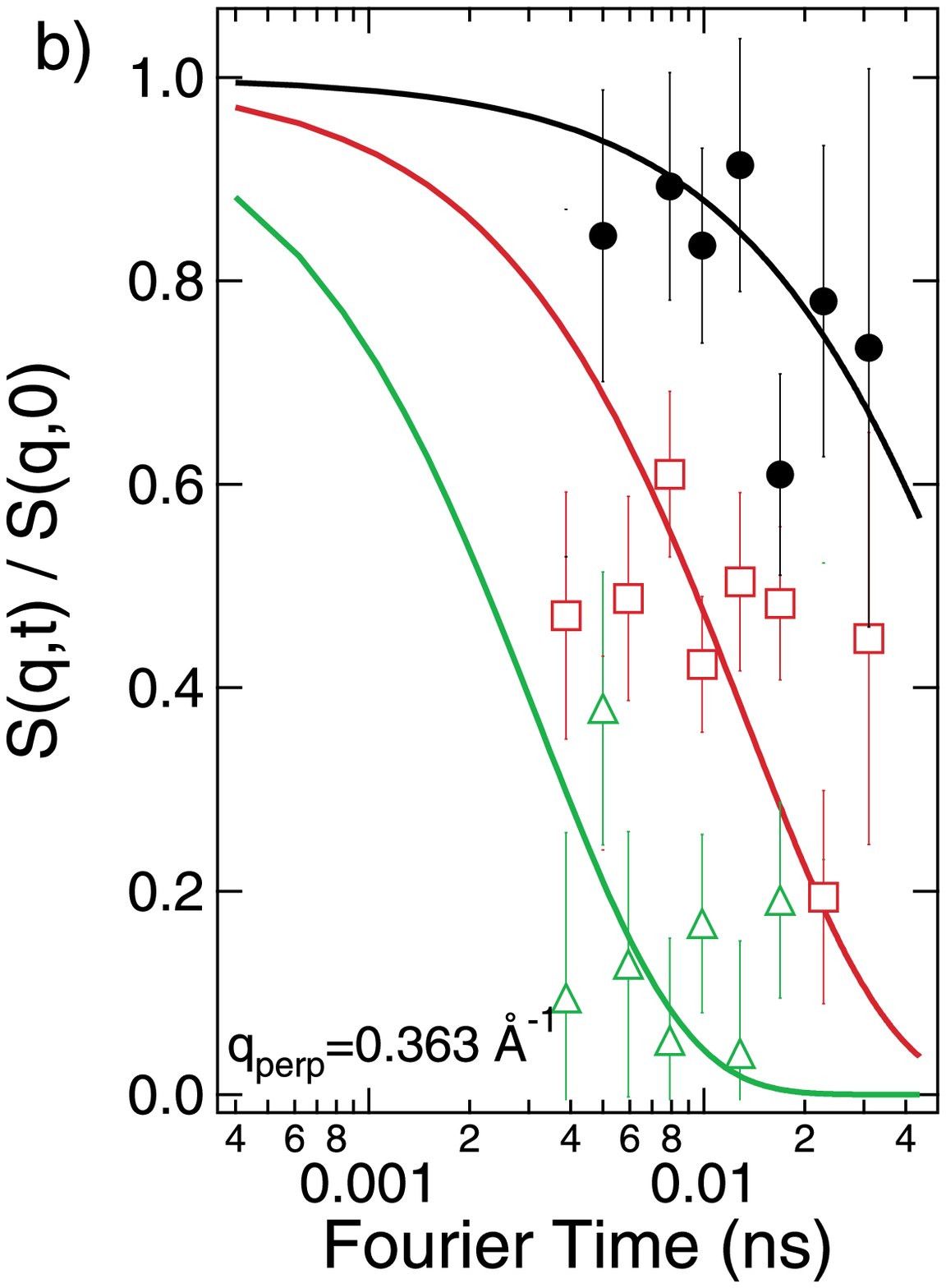}}}
\resizebox{0.35\textwidth}{!}{\rotatebox{0}{\includegraphics{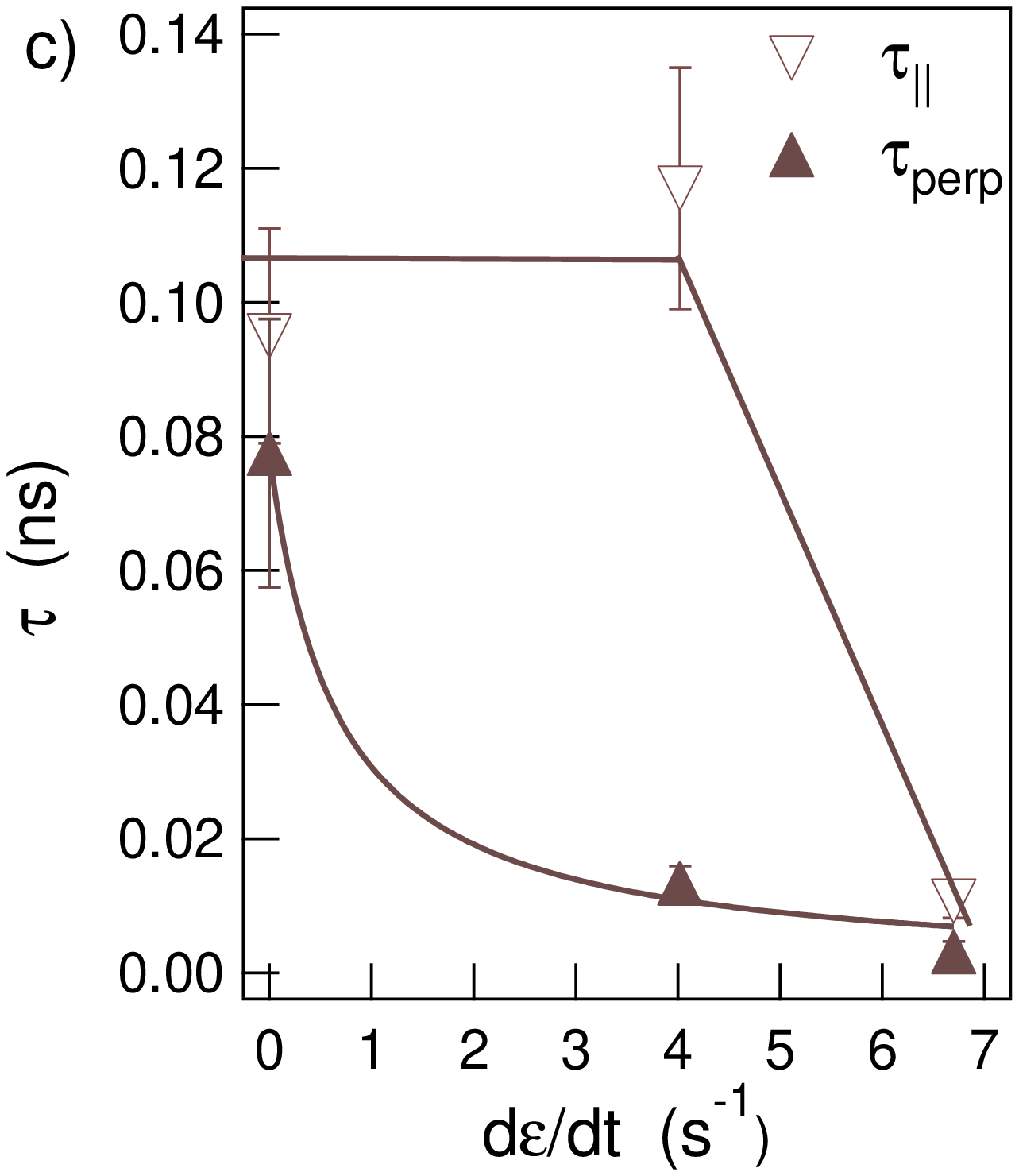}}}
\caption{(Color online) Intermediate scattering function
$S(q,t)/S(q,0)$ at $q=0.363~$\AA$^{-1}$ (a) for $q_{\|}$ and (b) for
$q_{\perp}$. Solid lines are fits after single exponential decays.
The resulting values for $\tau_{\|}$ and $\tau_{\perp}$ are plotted
in part (c).}\label{Graph_q abhaengige Dynamik.eps}
\end{figure*}

With an applied flow the signal of the elongated polymers appeared
as a modulation of the coherent scattering signal. When
investigating the different single detectors and different
individual length scales, only for $q$-values around the structural
correlation, $q_K$, the dynamics was found to be different from the
dynamics of the carrier liquid (within the experimental resolution).
The five detectors, which covered the q-range of the peak, were
summed up to increase the statistics. This time data were normalized
to D$_2$O at rest, i.e., divided by the measurement of the cell
filled with heavy water, only, and no flow applied, to cancel out
the dynamics of the pure carrier liquid. By normalizing to D$_2$O
instead of the elastic sample, larger error bars are expected.
$S(q,t)/S(q,0)=1$ then means pure water dynamics. Data for D$_2$O
with an applied flow were taken for comparison but did not show any
effect in the length and time scales and resolution accessible by
this experiment. Inelastic data for parallel and perpendicular flow
for $q_K=0.363~$\AA$^{-1}$ are shown in Fig.~\ref{Graph_q abhaengige
Dynamik.eps} (a) and (b). $S(q,t)/S(q,0)$ shows a decrease with
increasing time. %Because higher Fourier times were strongly
%contaminated by the incoherent water scattering, only about one
%decade in time could be used for the analysis. However,
The data points fall on (coherent) single exponential relaxation
steps. For $q_{\|}$, data for no-flow and Flow~1 coincide and for
the highest flow rate (Flow 2), the relaxation shifts to faster
times. The perpendicular component appears to be more sensitive to
an applied flow as the corresponding curves in Fig.~\ref{Graph_q
abhaengige Dynamik.eps} (b) shift to faster times with increasing
flow rate. The corresponding relaxation times $\tau_{\|}$ and
$\tau_{\perp}$ were determined from fits after
single exponential decays and are %given in Table~\ref{tauvalues},
%and also
plotted in Fig.~\ref{Graph_q abhaengige Dynamik.eps} (c). Note that
by normalizing the data to the dynamics of the carrier liquid, the
strong incoherent water scattering was eliminated and the time
scales determined stem from the contribution of the polymer chains,
only.
%\begin{table}\centering
%\begin{tabular}{c|c|c}
%$\dot\epsilon$ ($s^{-1}$) & $\tau_{\|}$ (ps) & $\tau_{\perp}
%(ps)$\\\hline\hline
%0 & 95 & 77.5\\
%4 & 117 & 13.4\\
%6.7 & 10.7 & 3.2\\
%\end{tabular} \caption[]{Values for the the relaxation times
%determined from fits to the curves in Fig.~\ref{Graph_q abhaengige
%Dynamik.eps} (a) and (b).}\label{tauvalues}
%\end{table}

%------------------------------------------------------
One important finding is the weak elastic peak
(Fig.~\ref{GraphDiffraction.eps}a) on the length scale of the Kuhn
segments. Such a short range order within the polymer chains would
easily have been overlooked in previous experiments, firstly because
it is a very weak feature, which only showed up by integrating over
24~h. And secondly, the q-value of 0.37 \AA$^{-1}$ is right between
the q-range, which is usually covered in small angle experiments,
and normal, wide angle diffraction and orders of magnitude weaker
than any structural feature in PEO. The prominent length scale at
which the peak is observed is 17~\AA, which is close, but slightly
larger than the tabulated value of 10~\AA. Note that the Kuhn length
typically is determined by means of rheological measurements, i.e.,
by a macroscopic technique, with an inherently large uncertainty. We
therefore tentatively interpret the short range dynamic order in PEO
as the experimentally observed Kuhn length in this system. But
systematic studies, also including different polymers and labeled
systems are needed to clarify this point.

The quality of the inelastic data in Fig.~\ref{Graph_q abhaengige
Dynamik.eps} is intensity limited, with corresponding large error
bars. The data can reasonably be described by single exponential
decays but can not exclude different models, such as stretched
exponentials or superpositions of several relaxation processes. The
time scales determined and in particular the observation that all
processes shift to faster times with an applied flow is robust and
can easily be seen in the data, despite the relatively large
uncertainties.

The properties of a flexible polymer chain in elongational flow were
studied analytically by Hofmann, Winkler and Reineker
\cite{Hofmann00}. The authors calculated the relaxation times for
the exponential decay $\tau_{\|}$ and $\tau_{\bot}$ for a given
normal mode of the underlying bead rod model to $\tau_{\bot} =
a/(b+\dot\epsilon/2)$
 for the fluctuation dynamics perpendicular to
the flow, and $\tau_{\|} = a/(b-\dot\epsilon)$ for the parallel
orientation ($a$ and $b$ are model parameters). They thus predicted
a decrease of the relaxation time for fluctuations perpendicular to
the flow, but an increase for the in-flow component. The decrease of
$\tau_{\bot}$ is observed in the data and the solid line in
Fig.~\ref{Graph_q abhaengige Dynamik.eps} (c) is a fit of the
theoretical prediction, which shows a very good agreement with the
data. The increase in the parallel direction ($\tau_{\|}$) was
clearly not observed. $\tau_{\|}$ is constant for no-flow and Flow 1
and even decreases with the strongest flow applied. One reason might
be that the model excludes any hydrodynamic interactions, which
might of great importance particularly for flexible polymers such as
PEO.

%------------------------------------------------------------------------------------
%\section{Conclusion\label{Conclusion}}
In conclusion, we found first experimental evidence for an
anisotropic dynamics on the monomer scale in semi diluted polymer
solution in an elongational flow. Diffraction data show a structural
correlation along the chains on a length scale of 17~\AA, which
might be related to the Kuhn length in this system. Only at
$q$-values close to the structural peak the dynamics of the polymers
is different from that of the carrier liquid. Relaxation times in
flow ($\tau_{\|}$) and perpendicular to the flow direction
($\tau_{\bot}$) have been determined. $\tau_{\bot}$ appears to be
very sensitive to an applied flow and can be described by an
analytical model, while the predicted increase for $\tau_{\|}$ could
not be observed. Inelastic neutron scattering thereby gives a unique
access to the microscopic dynamics of polymers in flow down to the
monomer scale, and the additional $q$-information allows to link the
observed relaxation rates to internal length scales. Future
experiments will address different polymer systems and also labeled
polymers.

{\bf Acknowledgement:} We thank M.~Schmidt, Univ.\ Saarbr\"ucken,
for the realization of the flow cells, and the ILL for the
allocation of beam time.

%------------------------------------------------------------------------------------
%References
%\bibliographystyle{unsrt}
\bibliography{PEO_08222008}

\begin{thebibliography}{14}
\expandafter\ifx\csname natexlab\endcsname\relax\def\natexlab#1{#1}\fi
\expandafter\ifx\csname bibnamefont\endcsname\relax
  \def\bibnamefont#1{#1}\fi
\expandafter\ifx\csname bibfnamefont\endcsname\relax
  \def\bibfnamefont#1{#1}\fi
\expandafter\ifx\csname citenamefont\endcsname\relax
  \def\citenamefont#1{#1}\fi
\expandafter\ifx\csname url\endcsname\relax
  \def\url#1{\texttt{#1}}\fi
\expandafter\ifx\csname urlprefix\endcsname\relax\def\urlprefix{URL }\fi
\providecommand{\bibinfo}[2]{#2}
\providecommand{\eprint}[2][]{\url{#2}}

\bibitem[{\citenamefont{Perkins et~al.}(1997)\citenamefont{Perkins, Smith, and
  Chu}}]{Perkins1997}
\bibinfo{author}{\bibfnamefont{T.~T.} \bibnamefont{Perkins}},
  \bibinfo{author}{\bibfnamefont{D.~E.} \bibnamefont{Smith}}, \bibnamefont{and}
  \bibinfo{author}{\bibfnamefont{S.}~\bibnamefont{Chu}},
  \bibinfo{journal}{Science} \textbf{\bibinfo{volume}{276}},
  \bibinfo{pages}{2016} (\bibinfo{year}{1997}).

\bibitem[{\citenamefont{Petrov et~al.}(2006)\citenamefont{Petrov, Ohrt,
  Winkler, and Schwille}}]{Schwille2006}
\bibinfo{author}{\bibfnamefont{E.}~\bibnamefont{Petrov}},
  \bibinfo{author}{\bibfnamefont{T.}~\bibnamefont{Ohrt}},
  \bibinfo{author}{\bibfnamefont{R.}~\bibnamefont{Winkler}}, \bibnamefont{and}
  \bibinfo{author}{\bibfnamefont{P.}~\bibnamefont{Schwille}},
  \bibinfo{journal}{Phys. Rev. Lett.} \textbf{\bibinfo{volume}{97}},
  \bibinfo{pages}{258101} (\bibinfo{year}{2006}).

\bibitem[{\citenamefont{Bird et~al.}(1987)\citenamefont{Bird, Armstrong, and
  Hassager}}]{Bird87}
\bibinfo{author}{\bibfnamefont{R.}~\bibnamefont{Bird}},
  \bibinfo{author}{\bibfnamefont{R.}~\bibnamefont{Armstrong}},
  \bibnamefont{and} \bibinfo{author}{\bibfnamefont{O.}~\bibnamefont{Hassager}},
  \emph{\bibinfo{title}{Dynamics of polymeric liquids}}
  (\bibinfo{publisher}{Wiley}, \bibinfo{address}{New York},
  \bibinfo{year}{1987}).

\bibitem[{\citenamefont{Larson and Magda}(2004)}]{Larson2003}
\bibinfo{author}{\bibfnamefont{R.~G.} \bibnamefont{Larson}} \bibnamefont{and}
  \bibinfo{author}{\bibfnamefont{J.~J.} \bibnamefont{Magda}},
  \bibinfo{journal}{Macromolecules} \textbf{\bibinfo{volume}{22}},
  \bibinfo{pages}{3004} (\bibinfo{year}{2004}).

\bibitem[{\citenamefont{Wagner et~al.}(2003)\citenamefont{Wagner, Amarouchene,
  Doyle, and Bonn}}]{Wagner2003}
\bibinfo{author}{\bibfnamefont{C.}~\bibnamefont{Wagner}},
  \bibinfo{author}{\bibfnamefont{Y.}~\bibnamefont{Amarouchene}},
  \bibinfo{author}{\bibfnamefont{P.}~\bibnamefont{Doyle}}, \bibnamefont{and}
  \bibinfo{author}{\bibfnamefont{D.}~\bibnamefont{Bonn}},
  \bibinfo{journal}{Eur. Phys. Lett.} \textbf{\bibinfo{volume}{64}},
  \bibinfo{pages}{823} (\bibinfo{year}{2003}).

\bibitem[{\citenamefont{Anna and McKinley}(2001)}]{Anna2001}
\bibinfo{author}{\bibfnamefont{S.~L.} \bibnamefont{Anna}} \bibnamefont{and}
  \bibinfo{author}{\bibfnamefont{G.~H.} \bibnamefont{McKinley}},
  \bibinfo{journal}{J. Rheol.} \textbf{\bibinfo{volume}{45}},
  \bibinfo{pages}{115} (\bibinfo{year}{2001}).

\bibitem[{\citenamefont{Stelter et~al.}(2002)\citenamefont{Stelter, Brenn,
  Yarin, Singh, and Durst}}]{Stelter2002}
\bibinfo{author}{\bibfnamefont{M.}~\bibnamefont{Stelter}},
  \bibinfo{author}{\bibfnamefont{G.}~\bibnamefont{Brenn}},
  \bibinfo{author}{\bibfnamefont{A.~L.} \bibnamefont{Yarin}},
  \bibinfo{author}{\bibfnamefont{R.~P.} \bibnamefont{Singh}}, \bibnamefont{and}
  \bibinfo{author}{\bibfnamefont{F.}~\bibnamefont{Durst}}, \bibinfo{journal}{J.
  Rheol.} \textbf{\bibinfo{volume}{46}}, \bibinfo{pages}{507}
  (\bibinfo{year}{2002}).

\bibitem[{\citenamefont{Bent et~al.}(2003)\citenamefont{Bent, Hutchings,
  Richards, Gough, Spares, Coates, Grillo, Harlen, Read, Graham
  et~al.}}]{Bent2003}
\bibinfo{author}{\bibfnamefont{J.}~\bibnamefont{Bent}},
  \bibinfo{author}{\bibfnamefont{L.~R.} \bibnamefont{Hutchings}},
  \bibinfo{author}{\bibfnamefont{R.~W.} \bibnamefont{Richards}},
  \bibinfo{author}{\bibfnamefont{T.}~\bibnamefont{Gough}},
  \bibinfo{author}{\bibfnamefont{R.}~\bibnamefont{Spares}},
  \bibinfo{author}{\bibfnamefont{P.~D.} \bibnamefont{Coates}},
  \bibinfo{author}{\bibfnamefont{I.}~\bibnamefont{Grillo}},
  \bibinfo{author}{\bibfnamefont{O.~G.} \bibnamefont{Harlen}},
  \bibinfo{author}{\bibfnamefont{D.~J.} \bibnamefont{Read}},
  \bibinfo{author}{\bibfnamefont{R.~S.} \bibnamefont{Graham}},
  \bibnamefont{et~al.}, \bibinfo{journal}{Science}
  \textbf{\bibinfo{volume}{301}}, \bibinfo{pages}{1691} (\bibinfo{year}{2003}).

\bibitem[{\citenamefont{Brandrup et~al.}(1999)\citenamefont{Brandrup, Immergut,
  and Grulke}}]{Brandrup1999}
\bibinfo{editor}{\bibfnamefont{J.}~\bibnamefont{Brandrup}},
  \bibinfo{editor}{\bibfnamefont{E.}~\bibnamefont{Immergut}}, \bibnamefont{and}
  \bibinfo{editor}{\bibfnamefont{E.}~\bibnamefont{Grulke}}, eds.,
  \emph{\bibinfo{title}{Polymer handbook}} (\bibinfo{publisher}{Wiley},
  \bibinfo{address}{New York}, \bibinfo{year}{1999}), \bibinfo{edition}{4th}
  ed.

\bibitem[{\citenamefont{Rodd et~al.}(2007)\citenamefont{Rodd, Cooper-White,
  Boger, and McKinley}}]{Rodd2007}
\bibinfo{author}{\bibfnamefont{L.}~\bibnamefont{Rodd}},
  \bibinfo{author}{\bibfnamefont{J.}~\bibnamefont{Cooper-White}},
  \bibinfo{author}{\bibfnamefont{D.}~\bibnamefont{Boger}}, \bibnamefont{and}
  \bibinfo{author}{\bibfnamefont{G.}~\bibnamefont{McKinley}},
  \bibinfo{journal}{J. Non-Newt. Fluid Mech.} \textbf{\bibinfo{volume}{143}},
  \bibinfo{pages}{170} (\bibinfo{year}{2007}).

\bibitem[{\citenamefont{Mezei}(1980)}]{Mezei:1980}
\bibinfo{editor}{\bibfnamefont{F.}~\bibnamefont{Mezei}}, ed.,
  \emph{\bibinfo{title}{Neutron Spin Echo}} (\bibinfo{publisher}{Springer},
  \bibinfo{address}{Berlin}, \bibinfo{year}{1980}).

\bibitem[{\citenamefont{Richter et~al.}(2005)\citenamefont{Richter,
  Monkenbusch, Arbe, and Colmenero}}]{Richter:2005}
\bibinfo{author}{\bibfnamefont{D.}~\bibnamefont{Richter}},
  \bibinfo{author}{\bibfnamefont{M.}~\bibnamefont{Monkenbusch}},
  \bibinfo{author}{\bibfnamefont{A.}~\bibnamefont{Arbe}}, \bibnamefont{and}
  \bibinfo{author}{\bibfnamefont{J.}~\bibnamefont{Colmenero}},
  \emph{\bibinfo{title}{Neutron spin echo in polymer systems}}
  (\bibinfo{publisher}{Springer}, \bibinfo{address}{Berlin/Heidelberg},
  \bibinfo{year}{2005}).

\bibitem[{\citenamefont{Dore and Teixeira}(1991)}]{Dore:1991}
\bibinfo{editor}{\bibfnamefont{J.~C.} \bibnamefont{Dore}} \bibnamefont{and}
  \bibinfo{editor}{\bibfnamefont{J.}~\bibnamefont{Teixeira}}, eds.,
  \emph{\bibinfo{title}{Hydrogen-Bonded Liquids}} (\bibinfo{publisher}{Kluwer},
  \bibinfo{address}{Dordrecht}, \bibinfo{year}{1991}).

\bibitem[{\citenamefont{Hofmann et~al.}(2000)\citenamefont{Hofmann, Winkler,
  and Reineker}}]{Hofmann00}
\bibinfo{author}{\bibfnamefont{T.}~\bibnamefont{Hofmann}},
  \bibinfo{author}{\bibfnamefont{R.~G.} \bibnamefont{Winkler}},
  \bibnamefont{and} \bibinfo{author}{\bibfnamefont{P.}~\bibnamefont{Reineker}},
  \bibinfo{journal}{Phys. Rev. E} \textbf{\bibinfo{volume}{61}},
  \bibinfo{pages}{2840} (\bibinfo{year}{2000}).

\end{thebibliography}

\end{document}